\documentclass[aps,prl,twocolumn,superscriptaddress,floatfix]{revtex4-1}
\usepackage{amsmath}
\usepackage{graphicx}
\usepackage{graphics}
\usepackage{hyperref}
\usepackage{physics}
\usepackage{color}

\begin{document}
\title{Interplay of non-Hermitian skin effects and Anderson localization in non-reciprocal quasiperiodic lattices}
\author{Hui Jiang}
\affiliation{Beijing National Laboratory for Condensed Matter Physics, Institute of Physics, Chinese Academy of Sciences, Beijing 100190, China}
\affiliation{School of Physical Sciences, University of Chinese Academy of Sciences, Beijing 100049, China}
\author{Li-Jun Lang}
\email{ljlang@scnu.edu.cn}
\affiliation{Guangdong Provincial Key Laboratory of Quantum Engineering and Quantum Materials, SPTE, South China Normal University, Guangzhou 510006, China}
\author{Chao Yang}
\affiliation{Beijing National Laboratory for Condensed Matter Physics, Institute of Physics, Chinese Academy of Sciences, Beijing 100190, China}
\affiliation{School of Physical Sciences, University of Chinese Academy of Sciences, Beijing 100049, China}
\author{Shi-Liang Zhu}
\affiliation{National Laboratory of Solid State Microstructures and School of Physics, Nanjing University, Nanjing 210093, China}
\affiliation{Guangdong Provincial Key Laboratory of Quantum Engineering and Quantum Materials, SPTE, South China Normal University, Guangzhou 510006, China}
\author{Shu Chen}
\email{schen@iphy.ac.cn}
\affiliation{Beijing National Laboratory for Condensed Matter Physics, Institute of Physics, Chinese Academy of Sciences, Beijing 100190, China}
\affiliation{School of Physical Sciences, University of Chinese Academy of Sciences, Beijing 100049, China}
\affiliation{The Yangtze River Delta Physics Research Center, Liyang, Jiangsu 213300, China}
\date{\today}

\begin{abstract}
Non-Hermiticity from non-reciprocal hoppings has been shown recently to demonstrate the non-Hermitian skin effect (NHSE) under open boundary conditions (OBCs). Here we study the interplay of this effect and the Anderson localization in a \textit{non-reciprocal} quasiperiodic lattice, dubbed non-reciprocal Aubry-Andr\'{e} model, and a \textit{rescaled} transition point is exactly proved. The non-reciprocity can induce not only the NHSE, but also the asymmetry in localized states with two Lyapunov exponents for both sides. Meanwhile, this transition is also topological, characterized by a winding number associated with the complex eigenenergies under periodic boundary conditions (PBCs), establishing a \textit{bulk-bulk} correspondence.  This interplay can be realized by an elaborately designed electronic circuit with only linear passive RLC devices instead of elusive non-reciprocal ones, where the transport of a continuous wave undergoes a transition between insulating and amplifying. This initiative scheme can be immediately applied in experiments to other non-reciprocal models, and will definitely inspires the study of interplay of NHSEs and more other quantum/topological phenomena.
\end{abstract}

\maketitle

%\section{Introduction}
Anderson localization (AL)  \cite{111124-3} is an old but everlasting research problem in condensed matters, which reveals a mechanism of insulation due to the destructive interference of multiple scattered waves induced by randomness \cite{140322-1,140327-1}. This fundamental phenomenon has been observed in experiments for electronic spins \cite{Feher59-1,Feher59-2}, light \cite{Wiersma97,Scheffold99,Schwartz07,Aegerter07}, microwave \cite{Dalichaouch91,Chabanov00,Pradhan00}, sound \cite{Weaver90}, and cold atoms \cite{Billy08,Roati08,Luschen18}. 
In one dimensional (1D) systems, it is well known that any infinitesimal disorder can localize all eigenstates \cite{111124-3,140322-1,140327-1}. However, it was found that relaxing the condition of randomness, the AL can also exist in quasiperiodic systems, e.g., Aubry-Andr\'{e} (AA) model \cite{131104-1}, but with a finite transition point.
This quasiperiodicity also has profound connection to topology \cite{110916-1,111216-1,130426-2,130524-1,130703-1}: The AA model can be mapped to the two dimensional Hofstadter model \cite{100903-1} with an external periodic parameter as a synthetic dimension, and thus realizes the famous Thouless pumping \cite{131116-1,111216-4P,160415-1,160415-2}.

On the other hand, non-Hermiticity \cite{110707-1} has been studied intensively for years with the aid of the fast development of the topological photonics \cite{161004-1,Ozawa18}; it exhibits rich phenomena without Hermitian counterparts, e.g., $\mathcal{PT}$ symmetry breaking \cite{Bender98,Guo09,Peng14}, exceptional points \cite{Zhen15,Ding16,Doppler16,Xu16,Midya18}, etc. Especially, the non-Hermitian topology is attracting special attention for the violation of the conventional bulk-boundary correspondence of Hermitian topological systems, and new ways of topological characterization are needed \cite{180720-9,Hu11,Esaki2011,Zhu14,171012-1,180720-8,Jin17,180720-7,Lieu18,180720-3,181022-1,Torres18,180720-6,181016-1,180720-4,181019-1,181221-2,181221-1,181127-1,181221-3,180924-1,190106-1,Harari18,Bandres18}. Besides the on-site gain/loss, non-reciprocal hoppings can also bring in non-Hermiticity \cite{180720-3,181022-1,Torres18,180720-6,181016-1,180720-4,181019-1,181221-2,181221-1,181127-1,181221-3} with exotic features, e.g., the topological non-Hermitian skin effect (NHSE) under open boundary conditions (OBCs), which is helpful to understand the breakdown of bulk-boundary correspondence.

Among references, effects of non-Hermiticity on AL have been studied in different contexts \cite{181017-2,181017-3,Shnerb98,Moiseyev01,Heinrichs01,181017-8,Longhi14,181017-6,181017-7,181017-1}, but the discussion on the interplay of NHSEs and the AL with accompanying topological transitions is still lacking. 
Thus, natural questions arise: What is the fate of the NHSE and its topology in the presence of quasiperiodic potentials, whether there is a
transition inherited from the well-known AL of the Hermitian AA model, and if yes, what is it like?
In this paper, we address the above questions in the AA model with non-reciprocal hoppings, dubbed the ``non-reciprocal AA model", and find the transition of NHSEs and AL under OBCs with an analytically proved \textit{rescaled} transition point. Affected by the non-reciprocity, besides the NHSE under OBCs, the localized states are asymmetric with respect to the localization center, characterized by \textit{two} Lyapunov exponents on both sides.
Meanwhile, this transition is topological, in the sense of the winding number associated with the complex eigenenergies under periodic boundary conditions (PBCs) \cite{181016-1}, which can well distinguish the different skin phases and the localized phase under OBCs, establishing a \textit{bulk-bulk} correspondence.
In the end, to demonstrate the interplay, an electronic circuit is elaborately proposed with \textit{only} linear passive RLC elements, which undoubtedly shows the phase transition through the transport of continuous waves between insulating and amplifying. Due to the lacking of experimental realizations of NHSEs, especially in electronic circuits \cite{170301-1,CHLee2018,181219-1,171211-1,Yu2018,190106-2,Hofmann18,Lu19}, our design is very practicable and can be immediately applied to other non-reciprocal models, and will definitely inspire the study of interplays of NHSEs and other quantum/topological phenomena.

\begin{figure}[tbp]
	\centering
	\includegraphics[width=1\columnwidth]{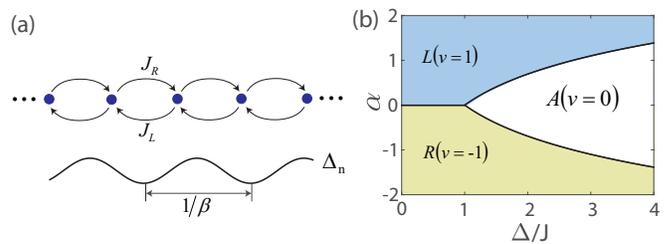}
	\caption{(a) Sketch of the non-reciprocal AA model.
		(b) Phase diagram. The phase boundaries are determined by $\Delta/J=e^{|\alpha|}$ and $\alpha=0$. Under OBCs, \{$L, R, A$\} represent the left-skin, right-skin, and Anderson localized phases, respectively. The winding number $\nu$ is defined in the text. Under PBCs, only regions $L$ and $R$ have imaginary eigenenergies. }\label{phase}
\end{figure}

\textit{Non-reciprocal AA model.}--The Hamiltonian of the non-reciprocal AA model [Fig. \ref{phase}(a)] reads
\begin{equation}\label{AAmodel1}
  \hat{H}=\sum_n(J_R\dyad{n+1}{n} +J_L\dyad{n}{n+1}+\Delta_n \dyad{n}{n}),
\end{equation}
where $J_{R(L)}$ is the right(left)-hopping amplitude, and $\Delta_n=2\Delta \cos(2\pi\beta n)$ is an on-site quasiperiodic potential with $\Delta$, without loss of generality, set positive and $\beta$ usually taken to be an irrational number, say, the inverse of the golden ratio $(\sqrt{5}-1)/2$ for infinite systems. For finite systems with site number $N=F_{n+1}$, where $F_n$ is $n$th Fibonacci number, because $\displaystyle\lim_{n\rightarrow\infty}F_n/F_{n+1}=(\sqrt{5}-1)/2$, we usually take the rational number $\beta=F_n/F_{n+1}$, preserving the quasiperiodicity.
For simplicity, we restrict the hoppings to be positive, which can be parameterized as $J_R=Je^{-\alpha}, J_L=Je^\alpha$ with $J>0$ and $\alpha$ both real, unless mentioned otherwise.
The non-reciprocity of hoppings ($\alpha\ne 0$) leads to the non-Hermiticity of the model, different from the non-Hermitian models based on the on-site gain/loss.

It is well known that, in the Hermitian case ($\alpha=0$), AL occurs at $\Delta/J=1$ for infinite systems due to the self-duality \cite{131104-1}:
The extended states for $\Delta/J<1$ become exponentially localized when $\Delta/J>1$ with the form $\ket{\psi}\propto \sum_n e^{-\eta|n-n_0|}\ket{n}$, where $n_0$ is the index of the localization center, and $\eta=\ln (\Delta/J)>0$ is the Lyapunov exponent, i.e., the inverse of the decaying length. 

Deviated from the Hermitian limit, the transition should be extended to the non-reciprocal case ($\alpha\ne 0$). 
To catch a glimpse of the non-reciprocity effect on the transition, we can quickly look into the two limits of the Hermitian case: 1) For the state fully localized at one site, i.e., $\Delta/J\rightarrow\infty$, because the sites are decoupled, the non-reciprocal hoppings have no effect on the state; 2) For the state extended through all sites, i.e.,
$\Delta/J\rightarrow 0$, under OBCs the non-reciprocal hoppings will accumulate the state to one boundary, i.e., the NHSE, depending on sgn$(\alpha)$ \cite{181022-1}. 
Apparently, at least under OBCs, the non-reciprocal AA model should undergo a transition between the skin phase and the localized phase. 

%\section{Interplay of NHSE and localization}
\textit{NHSE versus AL.-} To understand the AL in the non-reciprocal AA model,  Hamiltonian \eqref{AAmodel1} under OBCs can be rewritten in a biorthogonal basis as
$\hat{H}= \sum_{mn} h_{mn}\dyad{m}{n}=\sum_{mn} h'_{mn}\dyad{\tilde{m}_R}{\tilde{n}_L},$
where $\ket{\tilde{m}_R}\equiv e^{-\alpha m}\ket{m}$ and $\bra{\tilde{n}_L}\equiv\bra{n}e^{\alpha n}$ are the scaled basis in the right and left spaces, respectively, satisfying the biorthogonal condition $\braket{\tilde{n}_L}{\tilde{m}_R}=\delta_{mn}$.
Via this transformation, the non-Hermitian matrix $h$ becomes a Hermitian one,
\begin{equation}\label{hprime}
h'=\begin{pmatrix}
\Delta_1&J&&&\\
J&\Delta_2&J&&\\
&\ddots&\ddots&\ddots&\\
&&J&\Delta_{N-1}&J\\
&&&J&\Delta_N
\end{pmatrix},
\end{equation}
which is just the matrix representation of the Hermitian AA model with $J=\sqrt{J_L J_R}$ being the amplitude of the \textit{reciprocal} hoppings.
This transformation also reveals the fact that all eigenenergies of Hamiltonian \eqref{AAmodel1} are real, because $h$ and $h'$ are similar with the relation $h'=ShS^{-1}$, where $S=\text{diag} (e^{\alpha},e^{2\alpha},...,e^{N\alpha})$ is a similarity matrix with exponentially decaying diagonal entries.

As mentioned before, the Hermitian AA model represented by $h'$ undergoes AL at $\Delta/J=1$. Take $\psi'$ to be the eigenvector of $h'$. Mathematically, the right eigenvector of $h$ satisfies $\psi=S^{-1}\psi'$, which clearly shows how the non-reciprocity affects the state in the two phases of $h'$: For extended states, $S^{-1}$ exponentially accumulates the wave functions to one boundary, i.e., the NHSE; for localized states, the wave functions,
\begin{equation}\label{localization}
\psi_n \propto  
\left\{
\begin{array}{cc}
e^{-(\eta+\alpha)(n-n_0)},&n>n_0 \\
e^{-(\eta-\alpha)(n_0-n)}, &n<n_0 \\
\end{array}
\right.,
\end{equation}
manifest different decaying behaviors on both sides of the localization center with two Lyapunov exponents $\eta\pm\alpha$. These results are consistent with our previous limit analysis, reflecting the \textit{interplay} of the NHSE and the AL. According to Eq. \eqref{localization}, when $\eta \le |\alpha|$ delocalization occurs on one side and then skin modes emerge to the boundary on the same side, from which the boundary of skin/localized phases is given by 
\begin{equation}
\Delta/J=e^{|\alpha|}~~ \text{or}~~ \Delta/\max(J_L,J_R)=1.
\end{equation}
This transition is similar to the Hermitian case but determined by the larger hopping, which also determines to which skin the wave functions will accumulate after delocalization, and thus, the Hermitian case $(\alpha=0)$ separates the left-skin ($\alpha>0$) and right-skin  ($\alpha>0$) phases. 
Fig. \ref{phase}(b) shows the whole phase diagram.

\begin{figure}[tbp]
	\centering
	\includegraphics[width=1\columnwidth]{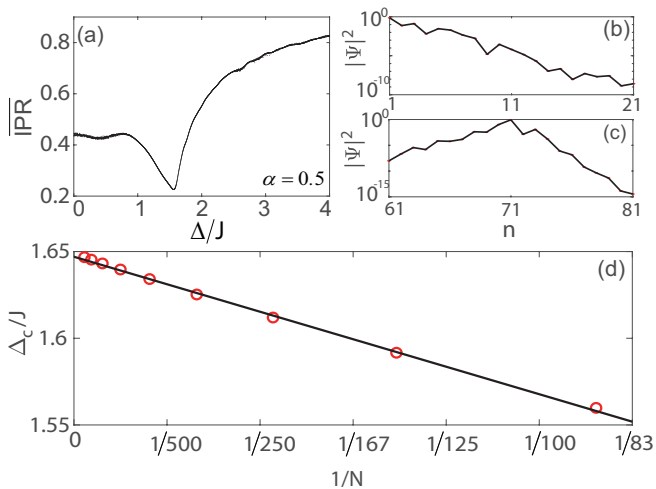}
	\caption{(a) $\overline{\text{IPR}}$ v.s. $\Delta/J$ for $\alpha=0.5$ under OBCs. The deep dive at $\sim1.56$ divides the skin and localized phases.
		The calculation is carried on with $N=89$ and $\beta=55/89$.
		(b,c) The profiles of the eigenstates of $10$th lowest $|E|$ in (a), showing the NHSE and the AL at $\Delta/J=0.5$ and $3$, respectively. 
		(d) Finite-size scaling analysis for the minimum $\overline{\text{IPR}}$, $\Delta_c/J$ (circles), of different lengths with the linear fitting (line), showing the asymptotic value $1.647\pm0.001$ when $N\rightarrow\infty$.
	}\label{OBC}
\end{figure}

As a demonstration, we calculate the averaged inverse participation ratios (IPRs) over all right eigenstates of $\hat{H}$ under OBCs,
\begin{equation}
\overline{\text{IPR}}=\frac{1}{N}\sum_{s=1}^{N}\text{IPR}_s=\frac{1}{N}\sum_{s=1}^{N}\frac{\sum_n\abs{\bra{n}\ket{\psi_s}}^4}{(\bra{\psi_s}\ket{\psi_s})^2},
\end{equation}
where $\ket{\psi_s}$ is the $s$th right eigenstate of $\hat{H}$. A state with $\text{IPR}=1$ is completely localized at a single site, while it is homogeneously distributed through all sites with $\text{IPR}=1/N$.
Different from the extended phase with small IPRs of the Hermitian case, the skin phase should have larger values due to its boundary-localization nature. Therefore, the transition point should correspond to the most extended case, i.e., the smallest $\overline{\text{IPR}}$.
As expected, a deep dive at $\sim1.56$ is found in Fig. \ref{OBC}(a), close to the theoretically predicted $e^{\alpha=0.5}\approx 1.65$ under consideration of the finite size effect, which is verified by the finite-size scaling analysis in Fig. \ref{OBC}(d). 
Figures. \ref{OBC}(b) and \ref{OBC}(c) typically show the skin mode, which is exponentially decaying from one boundary, and the asymmetrically localized mode with different decaying lengths on both sides, respectively.

\begin{figure}[tbp]
	\centering
	\includegraphics[width=1\columnwidth]{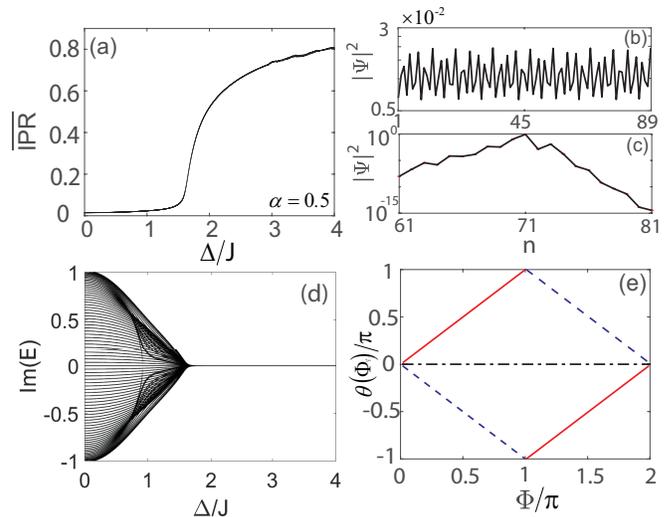}
	\caption{(a,b,c) The same setting as in Fig. \ref{OBC}(a-c) but under PBCs.
		(d) The imaginary parts of all eigenenergies in (a). 
		(e) $\theta(\Phi)$ for $\alpha=\Delta/J=0.5$ (solid red), $\alpha=-\Delta/J=-0.5$ (dashed blue), and $\alpha=0,\Delta/J=3$ (dash-dotted black), which correspond to $\nu=+1$, $-1$, and $0$, respectively, using the same $N$ and $\beta$.
}\label{PBC1}
\end{figure}

\textit{Periodic boundary conditions.}--Because of the breakdown of the conventional bulk-boundary correspondence, the behaviors under PBCs and OBCs should be much different. However, the insensitivity of the localized states to the boundaries hints that the onset of AL under both boundary conditions should be identical. This judgment is numerically verified in Fig. \ref{PBC1}(a): A steep rise of $\overline{\text{IPR}}$ around $e^{\alpha}$. Different from OBCs, the $\overline{\text{IPR}}$ keeps low prior to the transition due to the lacking of the localized skin modes [Fig. \ref{PBC1}(b)], while the localized states possess the same feature as OBCs [Fig. \ref{PBC1}(c)].

Another big difference is the presence of imaginary eigenenergies [Fig. \ref{PBC1}(d)]; the emergence of corner entries in $h$ invalidates the similarity to a Hermitian matrix. 
This feature is intimately related to the phase transition if we are reminded that the localized states are insensitive to the boundaries and thus have the real eigenenergies: The complexity-reality transition of the eigenenergies coincides with the AL.
Using this tie, we may establish a \textit{bulk-bulk} correspondence between systems under OBCs and PBCs through a winding number with respect to the complex eigenenergies.

%\section{Winding numbers}
\textit{Winding number.}--The conventional winding number cannot be used here because the chiral symmetry is broken by the on-site quasiperiodic potential \cite{180720-6,190106-1}.
Thus, we consider the ring chain with a magnetic flux $-\Phi$ penetrating through the center, yielding
\begin{eqnarray}\label{AAmodelphi}
\hat{H}(\Phi)&=&\hat{H}+J_R e^{ -i\Phi}\dyad{1}{N}+ J_L e^{ i\Phi}\dyad{N}{1}, 
\end{eqnarray}
and the winding number is defined as \cite{181016-1}
\begin{eqnarray}\label{winding1}
\nu=\frac{1}{2\pi i}\int_0^{2\pi} \text{d}\Phi\partial_{\Phi}\ln \det \hat{H}(\Phi) =\frac{1}{2\pi }\int_0^{2\pi}  \partial_\Phi\theta(\Phi) \text{d}\Phi,\notag \\
\end{eqnarray}
where $\theta(\Phi)$ is the argument of det$\hat{H}(\Phi)$.  Apparently, $\nu=0$ for the localized phase on account of the reality of the spectrum.

Figure \ref{PBC1}(e) show numerically how $\theta(\Phi)$ changes with $\Phi$ from $0$ to $2\pi$ in the three phases of Fig. \ref{phase}(b), and the corresponding winding numbers are obtained. 
The phase boundaries can alternatively be determined by analyzing the asymptotical behavior of det$H(\Phi)$ (See Supplemental Material).
As a result, the chirality of the winding number can exactly tell the left/right-skin phases ($\nu=\pm1$) and the localized phase ($\nu=0$) under OBCs.
Different from the conventional bulk-boundary correspondence, where edge states under OBCs can be predicted by a topological invariant defined under PBCs, here we establish a \textit{bulk-bulk} correspondence, where the behavior of bulk states under OBCs can be predicted by a topological invariant defined under PBCs.

\begin{figure}[tbp]
\centering
\includegraphics[width=1\columnwidth]{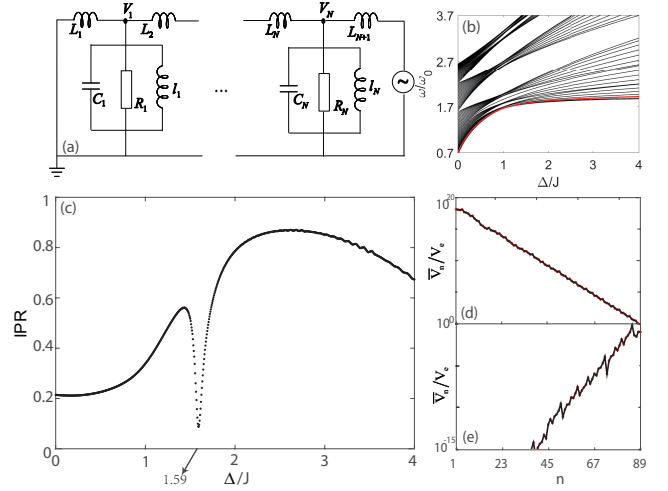}
\caption{(a) Schematic of the driven RLC electronic circuit with parameters defined in the text.
	(b) Intrinsic eigenfrequencies $\omega/\omega_0$ v.s. $\Delta/J$ for $\alpha=0.5$.
(c) IPR of $\overline{\mathcal{V}}$ v.s. $\Delta/J$ with driving frequencies indicated by the red curve in (b).
(d,e) Typical plots of $\overline{V}_n/V_e$ in (c) for $(\Delta/J,\tilde{\Omega})=(0.5,1.38)$ and $(3,1.88)$, respectively.
$N=89$ and $\beta=55/89$ are used.
}\label{dyn1}
\end{figure}

%\section{Electronic circuit's realization}
\textit{Electronic circuit's realization.}--We propose a driven RLC electronic circuit for the non-reciprocal AA model under OBCs, as shown in Fig. \ref{dyn1}(a), where inductors with inductances $L_n=Lg^{-n}$ and $l_n=Lg^{-n}[2\Delta(\cos2\pi\beta n+1)]^{-1}$, capacitors with capacitance $C_n=Cg^n$, and resistors with resistance $R_n=Rg^{-n}$ are all linear passive elements with positive free parameters, $L,C,R$, and $g$. 
The leftmost node is grounded for an open boundary while the other is connected to a voltage source of a continuous wave, $V_e(t) =V_e\sin(\Omega t)$ with driving frequency $\Omega$.
 
Without resistors, the intrinsic eigenfrequencies $\omega$ can be obtained by grounding the rightmost node instead of the source. Based on the Kirchhoff's current law, the corresponding eigenvalue equation reads,
\begin{equation}\label{circuitOBC}
  \begin{split}
   V_{n-1}+gV_{n+1}-\Delta_n V_n&=\left(f-\frac{\omega^2}{\omega_0^2} \right)V_n,
  \end{split}
\end{equation}
where $V_n$ is the amplitude of the voltage $V_n(t)$ on node $n$, $f=1+g+2\Delta$, and $\omega_0=1/\sqrt{LC}$. 
Rewritten in matrix form, $\mathcal{HV}=E\mathcal{V}$, where $\mathcal{V}=(\{V_n\})^T$ is a column vector and $E=f-\omega^2/\omega^2_0$ is the eigenvalue, $\mathcal{H}$ is just the matrix representation of the non-reciprocal AA model \eqref{AAmodel1} under OBCs with $J_L=g$ and $J_R=1$.
Notably, this classical circuit can only have real $E$, which is consistent with the previous proof.
Figure \ref{dyn1}(b) shows the intrinsic eigenfrequencies $\omega/\omega_0$ versus $\Delta/J$ with $J=\sqrt{g}$ and $\alpha=(\ln g)/2$.

When driving the system, the transport of continuous waves in different phases can be detected; the introduction of resistors, as seen in the following, is for system to quickly stabilize. The inhomogeneous equation with dimensionless parameters reads
\begin{equation}\label{drive}
   \frac{d^2}{d\tau^2}\mathcal{V}(\tau)+\gamma\frac{d}{d\tau}\mathcal{V}(\tau)-(\mathcal{H}-f)\mathcal{V}(\tau)=\mathcal{V}_e\sin\tilde{\Omega} \tau,
\end{equation}
where $\gamma=\frac{1}{R}\sqrt{\frac{L}{C}}>0$, $\tau=\omega_0 t$, and $\mathcal{V}_e=(0,...,0,V_e)^T$. The `$\sim$' over the frequency hereafter means the frequency is dimensionless in unit of $\omega_0$.
The solution is
\begin{equation}\label{dynbos}
\begin{split}
  \mathcal{V}(\tau)= &\sum_s \mathcal{V}_s\Big[e^{-{\gamma \tau/2}}(c_s\cos\lambda_s \tau+d_s\sin\lambda_s \tau)\\
        &\quad \quad +{\mathcal{W}}_s^T\mathcal{V}_e (a_s\cos \tilde{\Omega} \tau+b_s\sin\tilde{\Omega} \tau)\Big],
  \end{split}
\end{equation}
where $a_s = \frac{\gamma\tilde{\Omega}}{\gamma^2\tilde{\Omega}^2+(\tilde{\Omega}^2-\tilde{\omega}_s^2)^2}$, $b_s = \frac{\tilde{\Omega}^2-\tilde{\omega}_s^2}{\gamma^2\tilde{\Omega}^2+(\tilde{\Omega}^2-\tilde{\omega}_s^2)^2}$, $\lambda_s =\sqrt{\tilde{\omega}^2_s-\gamma^2 /2}$, and $(c_s,d_s)$ are coefficients determined by initial conditions. $\mathcal{V}_s$ and ${\mathcal{W}}_s^{T}$ are $s$th right and left eigenvectors of $\mathcal{H}$, respectively, satisfying $\mathcal{W}^T_s\mathcal{V}_{s'}=\delta_{ss'}$. Note that if $\mathcal{V}_s$ is accumulated to one boundary, $\mathcal{W}_s$ is to the other, because $\mathcal{W}_s$ is the right eigenvector of $\mathcal{H}^T$. Thus, to detect the \textit{left} skin modes, the source should be connected to the \textit{right} end for the possible large overlap $\mathcal{W}^T_s\mathcal{V}_e$. 
In Eq. \eqref{dynbos}, the first part in the square brackets is the general solution, which, due to the resistance, will decay  in a long time limit and thus, the effect of initial conditions can be ignored; the second part is one specific solution, which is stable, oscillating with the driving frequency. Moreover, if $\gamma\ll1$, the system is resonant when $\Omega\approx\omega_s$ with a large value of $a_s$ and vanishing $b_s$, unless the overlap $\mathcal{W}^T_s\mathcal{V}_e$ is zero, and the corresponding right eigenvector $\mathcal{V}_s$ can be picked out.

The IPR of the time-averaged voltage vector, $\overline{\mathcal{V}}= \frac{1}{T}\int_\tau^{\tau+T}|\mathcal{V}(\tau)|\text{d}\tau$ with $T=2\pi/\tilde{\Omega}$ in $\tau\rightarrow\infty$ limit, is shown in Fig. \ref{dyn1}(c), where a deep dive at $\sim1.59$ is close to the transition point.
Figures \ref{dyn1}(d) and \ref{dyn1}(e) plot the typical transports in both phases at $\alpha=0.5$: In the skin phase, due to the existence of left-skin modes, the continuous wave is resonantly transferred and accumulated to the left boundary; while in the localized phase, because of the small overlap $\mathcal{W}^T_s\mathcal{V}_e$, the wave is confined to the right boundary without resonance. 
If the input is from the left boundary, the existence of right-skin modes at $\alpha=-0.5$ will benefit the transport from left to the right.
This indicates that NHSEs can enhance the wave transport and may be useful in applications. This initiative realization of the non-reciprocity by circuits can be immediately applied to other non-reciprocal models, e.g., the non-reciprocal Su-Schrieffer-Heeger model \cite{180720-3,181022-1,180720-4,180720-6}.

%\section{Discussions}
\textit{Discussion and conclusion.}--The phase diagram in Fig. \ref{phase}(b) is obtained for positive hoppings. For general complex hoppings with arbitrary phases $\phi_{R(L)}$, an identical phase diagram is found numerically. Although no proper way to relate it to the positive-hopping case due to the effective net flux between each two nearest-neighbor sites, the special case satisfying $\phi_R+\phi_L=n\pi$ ($n\in$ integer) can be proved exactly by the duality.
We note that this transformation can map the non-reciprocal model to the AA model with complex on-site potentials, which, in the new basis, shares a similar AL but has no topological NHSEs.
The details can be seen in Supplemental Material.

For the circuit's realization, typically the element values can be taken as $L\sim$ mH, $C\sim$ pF, and $R\sim$ k$\Omega$, i.e., $\omega_0=1/\sqrt{LC}\sim$ kHz, which is accessible in usual circuit experiments \cite{170301-1,CHLee2018,181219-1,171211-1,Yu2018,190106-2,Hofmann18}. For typical non-reciprocal hoppings, say $\alpha=0.2$ and thus $g=e^{0.4}\approx 1.49$, the element values can still drop in almost the same orders for $N=10$ sites with $L_n\sim\mu$H to mH, $C_n\sim $ pF, and $R_n\sim$ k$\Omega$.

In summary, we have revealed the interplay of NHSEs and AL in the non-reciprocal AA model with accompanying topologies, and obtained analytically the exact phase diagram. Especially, an elegant experimental scheme with electronic circuits has been proposed, demonstrating a transport transition from insulating to amplifying.

\begin{acknowledgments}
	SC was supported by the NSFC (Grants No. 11425419) and the NKRDP of China (Grants No. 2016YFA0300600 and No. 2016YFA0302104). LJL was supported by the startup funding from SCNU. SLZ was supported by the NSFC (Grants No. 91636218 and U1801661) and the NKRDP of China (Grant No. 2016YFA0301803).
\end{acknowledgments}

% references------------------------------------------------------------------

%\bibliography{ref}

\begin{thebibliography}{10}
	
	\bibitem{111124-3}
	P.~W. Anderson.
	\newblock Absence of diffusion in certain random lattices.
	\newblock {\em Phys. Rev.}, 109:1492--1505, Mar 1958.
	
	\bibitem{140322-1}
	Elihu Abrahams, editor.
	\newblock {\em 50 Years of Anderson Localization}.
	\newblock World Scientific, 1st edition, 2010.
	
	\bibitem{140327-1}
	Patrick~A. Lee and T.~V. Ramakrishnan.
	\newblock Disordered electronic systems.
	\newblock {\em Rev. Mod. Phys.}, 57:287--337, Apr 1985.
	
	\bibitem{Feher59-1}
	G.~Feher.
	\newblock Electron spin resonance experiments on donors in silicon. i.
	electronic structure of donors by the electron nuclear double resonance
	technique.
	\newblock {\em Phys. Rev.}, 114:1219--1244, Jun 1959.
	
	\bibitem{Feher59-2}
	G.~Feher and E.~A. Gere.
	\newblock Electron spin resonance experiments on donors in silicon. ii.
	electron spin relaxation effects.
	\newblock {\em Phys. Rev.}, 114:1245--1256, Jun 1959.
	
	\bibitem{Wiersma97}
	Diederik~S. Wiersma, Paolo Bartolini, Ad~Lagendijk, and Roberto Righini.
	\newblock Localization of light in a disordered medium.
	\newblock {\em Nature}, 390:671--, December 1997.
	
	\bibitem{Scheffold99}
	Frank Scheffold, Ralf Lenke, Ralf Tweer, and Georg Maret.
	\newblock Localization or classical diffusion of light?
	\newblock {\em Nature}, 398:206--, March 1999.
	
	\bibitem{Schwartz07}
	Tal Schwartz, Guy Bartal, Shmuel Fishman, and Mordechai Segev.
	\newblock Transport and anderson localization in disordered two-dimensional
	photonic lattices.
	\newblock {\em Nature}, 446:52--, March 2007.
	
	\bibitem{Aegerter07}
	Christof~M. Aegerter, Martin St\"{o}rzer, Susanne Fiebig, Wolfgang B\"{u}hrer,
	and Georg Maret.
	\newblock Observation of anderson localization of light in three dimensions.
	\newblock {\em J. Opt. Soc. Am. A}, 24(10):A23--A27, Oct 2007.
	
	\bibitem{Dalichaouch91}
	Rachida Dalichaouch, J.~P. Armstrong, S.~Schultz, P.~M. Platzman, and S.~L.
	McCall.
	\newblock Microwave localization by two-dimensional random scattering.
	\newblock {\em Nature}, 354:53--, November 1991.
	
	\bibitem{Chabanov00}
	A.~A. Chabanov, M.~Stoytchev, and A.~Z. Genack.
	\newblock Statistical signatures of photon localization.
	\newblock {\em Nature}, 404:850--, April 2000.
	
	\bibitem{Pradhan00}
	Prabhakar Pradhan and S.~Sridhar.
	\newblock Correlations due to localization in quantum eigenfunctions of
	disordered microwave cavities.
	\newblock {\em Phys. Rev. Lett.}, 85:2360--2363, Sep 2000.
	
	\bibitem{Weaver90}
	Richard~L Weaver.
	\newblock Anderson localization of ultrasound.
	\newblock {\em Wave Motion}, 12(2):129--142, 1 1990.
	
	\bibitem{Billy08}
	Juliette Billy, Vincent Josse, Zhanchun Zuo, Alain Bernard, Ben Hambrecht,
	Pierre Lugan, David Clement, Laurent Sanchez-Palencia, Philippe Bouyer, and
	Alain Aspect.
	\newblock Direct observation of anderson localization of matter waves in a
	controlled disorder.
	\newblock {\em Nature}, 453:891--, June 2008.
	
	\bibitem{Roati08}
	Giacomo Roati, Chiara D'Errico, Leonardo Fallani, Marco Fattori, Chiara Fort,
	Matteo Zaccanti, Giovanni Modugno, Michele Modugno, and Massimo Inguscio.
	\newblock Anderson localization of a non-interacting bose-einstein condensate.
	\newblock {\em Nature}, 453(7197):895--U36, JUN 12 2008.
	
	\bibitem{Luschen18}
	Henrik~P. L\"uschen, Sebastian Scherg, Thomas Kohlert, Michael Schreiber,
	Pranjal Bordia, Xiao Li, S.~Das~Sarma, and Immanuel Bloch.
	\newblock Single-particle mobility edge in a one-dimensional quasiperiodic
	optical lattice.
	\newblock {\em Phys. Rev. Lett.}, 120:160404, Apr 2018.
	
	\bibitem{131104-1}
	Serge Aubry and Gilles Andr\'{e}.
	\newblock Analyticity breaking and anderson localization in incommensurate
	lattices.
	\newblock {\em Ann. Israel. Phys. Soc.}, 3:133, 1980.
	
	\bibitem{110916-1}
	Li-Jun Lang, Xiaoming Cai, and Shu Chen.
	\newblock Edge states and topological phases in one-dimensional optical
	superlattices.
	\newblock {\em Phys. Rev. Lett.}, 108:220401, May 2012.
	
	\bibitem{111216-1}
	Feng Mei, Shi-Liang Zhu, Zhi-Ming Zhang, C.~H. Oh, and N.~Goldman.
	\newblock Simulating ${Z}_{2}$ topological insulators with cold atoms in a
	one-dimensional optical lattice.
	\newblock {\em Phys. Rev. A}, 85:013638, Jan 2012.
	
	\bibitem{130426-2}
	Xiaoming Cai, Li-Jun Lang, Shu Chen, and Yupeng Wang.
	\newblock Topological superconductor to anderson localization transition in
	one-dimensional incommensurate lattices.
	\newblock {\em Phys. Rev. Lett.}, 110:176403, Apr 2013.
	
	\bibitem{130524-1}
	Zhihao Xu, Linhu Li, and Shu Chen.
	\newblock Fractional topological states of dipolar fermions in one-dimensional
	optical superlattices.
	\newblock {\em Phys. Rev. Lett.}, 110:215301, May 2013.
	
	\bibitem{130703-1}
	Shi-Liang Zhu, Z.-D. Wang, Y.-H. Chan, and L.-M. Duan.
	\newblock Topological bose-mott insulators in a one-dimensional optical
	superlattice.
	\newblock {\em Phys. Rev. Lett.}, 110:075303, Feb 2013.
	
	\bibitem{100903-1}
	Douglas~R. Hofstadter.
	\newblock Energy levels and wave functions of bloch electrons in rational and
	irrational magnetic fields.
	\newblock {\em Phys. Rev. B}, 14(6):2239--2249, Sep 1976.
	
	\bibitem{131116-1}
	D.~J. Thouless.
	\newblock Quantization of particle transport.
	\newblock {\em Phys. Rev. B}, 27:6083--6087, May 1983.
	
	\bibitem{111216-4P}
	Yaacov~E. Kraus, Yoav Lahini, Zohar Ringel, Mor Verbin, and Oded Zilberberg.
	\newblock Topological states and adiabatic pumping in quasicrystals.
	\newblock {\em Phys. Rev. Lett.}, 109:106402, Sep 2012.
	
	\bibitem{160415-1}
	M.~Lohse, C.~Schweizer, O.~Zilberberg, M.~Aidelsburger, and I.~Bloch.
	\newblock A thouless quantum pump with ultracold bosonic atoms in an optical
	superlattice.
	\newblock {\em Nat Phys}, 12(4):350--354, April 2016.
	
	\bibitem{160415-2}
	Shuta Nakajima, Takafumi Tomita, Shintaro Taie, Tomohiro Ichinose, Hideki
	Ozawa, Lei Wang, Matthias Troyer, and Yoshiro Takahashi.
	\newblock Topological thouless pumping of ultracold fermions.
	\newblock {\em Nat Phys}, 12(4):296--300, April 2016.
	
	\bibitem{110707-1}
	Nimrod Moiseyev.
	\newblock {\em Non-Hermitian Quantum Mechanics}.
	\newblock CAMBRIDGE UNIVERSITY PRESS, 1st edition, 2011.
	
	\bibitem{161004-1}
	Ling Lu, John~D. Joannopoulos, and Marin Soljacic.
	\newblock Topological photonics.
	\newblock {\em Nat Photon}, 8(11):821--829, November 2014.
	
	\bibitem{Ozawa18}
	T.~Ozawa, H.~M. Price, A.~Amo, N.~Goldman, M.~Hafezi, L.~Lu, M.~Rechtsman,
	D.~Schuster, J.~Simon, O.~Zilberberg, and I.~Carusotto.
	\newblock Topological photonics.
	\newblock {\em arXiv:1802.04173}, 2018.
	
	\bibitem{Bender98}
	Carl~M. Bender and Stefan Boettcher.
	\newblock Real spectra in non-hermitian hamiltonians having $pt$ symmetry.
	\newblock {\em Phys. Rev. Lett.}, 80(24):5243--5246, 15 Jun 1998.
	
	\bibitem{Guo09}
	A.~Guo, G.~J. Salamo, D.~Duchesne, R.~Morandotti, M.~Volatier-Ravat, V.~Aimez,
	G.~A. Siviloglou, and D.~N. Christodoulides.
	\newblock Observation of $\mathcal{P}\mathcal{T}$-symmetry breaking in complex
	optical potentials.
	\newblock {\em Phys. Rev. Lett.}, 103:093902, Aug 2009.
	
	\bibitem{Peng14}
	Bo~Peng, Sahin~Kaya Ozdemir, Fuchuan Lei, Faraz Monifi, Mariagiovanna
	Gianfreda, Gui~Lu Long, Shanhui Fan, Franco Nori, Carl~M. Bender, and Lan
	Yang.
	\newblock Parity-time-symmetric whispering-gallery microcavities.
	\newblock {\em Nature Physics}, 10:394--, April 2014.
	
	\bibitem{Zhen15}
	Bo~Zhen, Chia~Wei Hsu, Yuichi Igarashi, Ling Lu, Ido Kaminer, Adi Pick,
	Song-Liang Chua, John~D. Joannopoulos, and Marin Soljacic.
	\newblock Spawning rings of exceptional points out of dirac cones.
	\newblock {\em Nature}, 525:354--, September 2015.
	
	\bibitem{Ding16}
	Kun Ding, Guancong Ma, Meng Xiao, Z.~Q. Zhang, and C.~T. Chan.
	\newblock Emergence, coalescence, and topological properties of multiple
	exceptional points and their experimental realization.
	\newblock {\em Phys. Rev. X}, 6:021007, Apr 2016.
	
	\bibitem{Doppler16}
	Jorg Doppler, Alexei~A. Mailybaev, Julian Bohm, Ulrich Kuhl, Adrian Girschik,
	Florian Libisch, Thomas~J. Milburn, Peter Rabl, Nimrod Moiseyev, and Stefan
	Rotter.
	\newblock Dynamically encircling an exceptional point for asymmetric mode
	switching.
	\newblock {\em Nature}, 537:76--, July 2016.
	
	\bibitem{Xu16}
	H.~Xu, D.~Mason, Luyao Jiang, and J.~G.~E. Harris.
	\newblock Topological energy transfer in an optomechanical system with
	exceptional points.
	\newblock {\em Nature}, 537:80--, July 2016.
	
	\bibitem{Midya18}
	Bikashkali Midya, Han Zhao, and Liang Feng.
	\newblock Non-hermitian photonics promises exceptional topology of light.
	\newblock {\em Nature Communications}, 9(1):2674--, 2018.
	
	\bibitem{180720-9}
	M.~S. Rudner and L.~S. Levitov.
	\newblock Topological transition in a non-hermitian quantum walk.
	\newblock {\em Phys. Rev. Lett.}, 102:065703, 2009.
	
	\bibitem{Hu11}
	Yi~Chen Hu and Taylor~L. Hughes.
	\newblock Absence of topological insulator phases in non-hermitian
	$pt$-symmetric hamiltonians.
	\newblock {\em Phys. Rev. B}, 84:153101, Oct 2011.
	
	\bibitem{Esaki2011}
	Kenta Esaki, Masatoshi Sato, Kazuki Hasebe, and Mahito Kohmoto.
	\newblock Edge states and topological phases in non-hermitian systems.
	\newblock {\em Phys. Rev. B}, 84:205128, Nov 2011.
	
	\bibitem{Zhu14}
	Baogang Zhu, Rong L\"u, and Shu Chen.
	\newblock $\mathcal{PT}$ symmetry in the non-hermitian su-schrieffer-heeger
	model with complex boundary potentials.
	\newblock {\em Phys. Rev. A}, 89:062102, Jun 2014.
	
	\bibitem{171012-1}
	Tony~E. Lee.
	\newblock Anomalous edge state in a non-hermitian lattice.
	\newblock {\em Phys. Rev. Lett.}, 116:133903, Apr 2016.
	
	\bibitem{180720-8}
	D.~Leykam, K.~Y. Bliokh, C.~Huang, Y.~D. Chong, and F.~Nori.
	\newblock Edge modes, degeneracies, and topological numbers in non-hermitian
	systems.
	\newblock {\em Phys. Rev. Lett.}, 118:040401, 2017.
	
	\bibitem{Jin17}
	L.~Jin.
	\newblock Topological phases and edge states in a non-hermitian trimerized
	optical lattice.
	\newblock {\em Phys. Rev. A}, 96:032103, Sep 2017.
	
	\bibitem{180720-7}
	Huitao Shen, Bo~Zhen, and Liang Fu.
	\newblock Topological band theory for non-hermitian hamiltonians.
	\newblock {\em Phys. Rev. Lett.}, 120:146402, Apr 2018.
	
	\bibitem{Lieu18}
	Simon Lieu.
	\newblock Topological phases in the non-hermitian su-schrieffer-heeger model.
	\newblock {\em Phys. Rev. B}, 97:045106, Jan 2018.
	
	\bibitem{180720-3}
	Chuanhao Yin, Hui Jiang, Linhu Li, Rong L\"u, and Shu Chen.
	\newblock Geometrical meaning of winding number and its characterization of
	topological phases in one-dimensional chiral non-hermitian systems.
	\newblock {\em Phys. Rev. A}, 97:052115, May 2018.
	
	\bibitem{181022-1}
	Shunyu Yao and Zhong Wang.
	\newblock Edge states and topological invariants of non-hermitian systems.
	\newblock {\em Phys. Rev. Lett.}, 121:086803, Aug 2018.
	
	\bibitem{Torres18}
	V.~M. Martinez~Alvarez, J.~E. Barrios~Vargas, and L.~E.~F. Foa~Torres.
	\newblock Non-hermitian robust edge states in one dimension: Anomalous
	localization and eigenspace condensation at exceptional points.
	\newblock {\em Phys. Rev. B}, 97:121401, Mar 2018.
	
	\bibitem{180720-6}
	Ye~Xiong.
	\newblock Why does bulk boundary correspondence fail in some non-hermitian
	topological models.
	\newblock {\em Journal of Physics Communications}, 2(3):035043, 2018.
	
	\bibitem{181016-1}
	Zongping Gong, Yuto Ashida, Kohei Kawabata, Kazuaki Takasan, Sho Higashikawa,
	and Masahito Ueda.
	\newblock Topological phases of non-hermitian systems.
	\newblock {\em Phys. Rev. X}, 8:031079, Sep 2018.
	
	\bibitem{180720-4}
	Flore~K. Kunst, Elisabet Edvardsson, Jan~Carl Budich, and Emil~J. Bergholtz.
	\newblock Biorthogonal bulk-boundary correspondence in non-hermitian systems.
	\newblock {\em Phys. Rev. Lett.}, 121:026808, Jul 2018.
	
	\bibitem{181019-1}
	Shunyu Yao, Fei Song, and Zhong Wang.
	\newblock Non-hermitian chern bands.
	\newblock {\em Phys. Rev. Lett.}, 121:136802, Sep 2018.
	
	\bibitem{181221-2}
	L.~Jin and Z.~Song.
	\newblock Bulk-boundary correspondence in non-hermitian systems.
	\newblock {\em arXiv:1809.03139}, 2018.
	
	\bibitem{181221-1}
	Ching~Hua Lee and Ronny Thomale.
	\newblock Anatomy of skin modes and topology in non-hermitian systems.
	\newblock {\em arXiv:1809.02125}, 2018.
	
	\bibitem{181127-1}
	Kohei Kawabata, Ken Shiozaki, and Masahito Ueda.
	\newblock Anomalous helical edge states in a non-hermitian chern insulator.
	\newblock {\em Phys. Rev. B}, 98:165148, Oct 2018.
	
	\bibitem{181221-3}
	Ching~Hua Lee, Linhu Li, and Jiangbin Gong.
	\newblock Hybrid higher-order skin-topological modes in non-reciprocal systems.
	\newblock {\em arXiv: 1810.11824}, 2018.
	
	\bibitem{180924-1}
	Li-Jun Lang, You Wang, Hailong Wang, and Y.~D. Chong.
	\newblock Effects of non-hermiticity on su-schrieffer-heeger defect states.
	\newblock {\em Phys. Rev. B}, 98:094307, Sep 2018.
	
	\bibitem{190106-1}
	Hui Jiang, Chao Yang, and Shu Chen.
	\newblock Topological invariants and phase diagrams for one-dimensional
	two-band non-hermitian systems without chiral symmetry.
	\newblock {\em Phys. Rev. A}, 98:052116, Nov 2018.
	
	\bibitem{Harari18}
	G.~Harari, M.~A. Bandres, Y.~Lumer, M.~C. Rechtsman, Y.~D. Chong,
	M.~Khajavikhan, D.~N. Christodoulides, and M.~Segev.
	\newblock Topological insulator laser: Theory.
	\newblock {\em Science}, 359:eaar4003, 2018.
	
	\bibitem{Bandres18}
	M.~A. Bandres, S.~Wittek, G.~Harari, M.~Parto, J.~Ren, M.~Segev, D.~N.
	Christodoulides, and M.~Khajavikhan.
	\newblock Topological insulator laser: Experiments.
	\newblock {\em Science}, 359:eaar4005, 2018.
	
	\bibitem{181017-2}
	Naomichi Hatano and David~R. Nelson.
	\newblock Localization transitions in non-hermitian quantum mechanics.
	\newblock {\em Phys. Rev. Lett.}, 77:570--573, Jul 1996.
	
	\bibitem{181017-3}
	Naomichi Hatano and David~R. Nelson.
	\newblock Vortex pinning and non-hermitian quantum mechanics.
	\newblock {\em Phys. Rev. B}, 56:8651--8673, Oct 1997.
	
	\bibitem{Shnerb98}
	Nadav~M. Shnerb and David~R. Nelson.
	\newblock Winding numbers, complex currents, and non-hermitian localization.
	\newblock {\em Phys. Rev. Lett.}, 80:5172--5175, Jun 1998.
	
	\bibitem{Moiseyev01}
	Nimrod Moiseyev and Markus Gl\"uck.
	\newblock Non-hermitian delocalization from hermitian hamiltonians.
	\newblock {\em Phys. Rev. E}, 63:041103, Mar 2001.
	
	\bibitem{Heinrichs01}
	J.~Heinrichs.
	\newblock Eigenvalues in the non-hermitian anderson model.
	\newblock {\em Phys. Rev. B}, 63:165108, Apr 2001.
	
	\bibitem{181017-8}
	Amin Jazaeri and Indubala~I. Satija.
	\newblock Localization transition in incommensurate non-hermitian systems.
	\newblock {\em Phys. Rev. E}, 63:036222, Feb 2001.
	
	\bibitem{Longhi14}
	Stefano Longhi.
	\newblock $\mathcal {PT}$-symmetric optical superlattices.
	\newblock {\em Journal of Physics A: Mathematical and Theoretical},
	47(16):165302, 2014.
	
	\bibitem{181017-6}
	C.~Yuce.
	\newblock Pt symmetric aubry-andr\'e model.
	\newblock {\em Physics Letters A}, 378(30):2024 -- 2028, 2014.
	
	\bibitem{181017-7}
	Charles~H. Liang, Derek~D. Scott, and Yogesh~N. Joglekar.
	\newblock $\mathcal{PT}$ restoration via increased loss and gain in the
	$\mathcal{PT}$-symmetric aubry-andr\'e model.
	\newblock {\em Phys. Rev. A}, 89:030102, Mar 2014.
	
	\bibitem{181017-1}
	Qi-Bo Zeng, Shu Chen, and Rong L\"u.
	\newblock Anderson localization in the non-hermitian aubry-andr\'e-harper model
	with physical gain and loss.
	\newblock {\em Phys. Rev. A}, 95:062118, Jun 2017.
	
	\bibitem{170301-1}
	Jia Ningyuan, Clai Owens, Ariel Sommer, David Schuster, and Jonathan Simon.
	\newblock Time- and site-resolved dynamics in a topological circuit.
	\newblock {\em Phys. Rev. X}, 5:021031, Jun 2015.
	
	\bibitem{CHLee2018}
	Ching~Hua Lee, Stefan Imhof, Christian Berger, Florian Bayer, Johannes Brehm,
	Laurens~W. Molenkamp, Tobias Kiessling, and Ronny Thomale.
	\newblock Topolectrical circuits.
	\newblock {\em Communications Physics}, 1(1):39--, 2018.
	
	\bibitem{181219-1}
	Motohiko Ezawa.
	\newblock Higher-order topological electric circuits and topological corner
	resonance on the breathing kagome and pyrochlore lattices.
	\newblock {\em Phys. Rev. B}, 98:201402, Nov 2018.
	
	\bibitem{171211-1}
	Yuan Li, Yong Sun, Weiwei Zhu, Zhiwei Guo, Jun Jiang, Toshikaze Kariyado, Hong
	Chen, and Xiao Hu.
	\newblock Topological lc-circuits based on microstrips and observation of
	electromagnetic modes with orbital angular momentum.
	\newblock {\em Nature Communications}, 9(1):4598--, 2018.
	
	\bibitem{Yu2018}
	Kaifa Luo, Rui Yu, and Hongming Weng.
	\newblock Topological nodal states in circuit lattice.
	\newblock {\em Research}, 2018:10, 2018.
	
	\bibitem{190106-2}
	You Wang, Li-Jun Lang, Ching~Hua Lee, Baile Zhang, and Y~D.~Chong.
	\newblock Topologically enhanced harmonic generation in a nonlinear
	transmission line metamaterial.
	\newblock {\em arXiv:1807.11163}, 07 2018.
	
	\bibitem{Hofmann18}
	Tobias Hofmann, Tobias Helbig, Ching~Hua Lee, Martin Greiter, and Ronny
	Thomale.
	\newblock Chiral voltage propagation and calibration in a topolectrical chern
	circuit.
	\newblock {\em arXiv:1809.08687}, 2018.
	
	\bibitem{Lu19}
	Yuehui Lu, Ningyuan Jia, Lin Su, Clai Owens, Gediminas
	Juzeli\ifmmode~\bar{u}\else \={u}\fi{}nas, David~I. Schuster, and Jonathan
	Simon.
	\newblock Probing the berry curvature and fermi arcs of a weyl circuit.
	\newblock {\em Phys. Rev. B}, 99:020302, Jan 2019.
	
\end{thebibliography}
%\bibliographystyle{unsrt} %{apsrev4-1} 

%---------------------------------------------------------------------------------------------------------------------------------------------------------------------------
\appendix

In this Supplemental Material, we present the duality of the nonreciprocal AA model, the calculation of the winding number, and the discussion of the general case with complex hoppings.

\section{Duality}\label{duality}
That the non-reciprocal AA model can be transformed to the AA model with a complex on-site potential, i.e., the duality, can work in two cases:
1) Under PBCs with $\beta=p/N$, where $p\in$ integer; 2) Under OBCs with $N\rightarrow\infty$, because these two cases can ensure that the transformed $k$-space is closed by the following Fourier transform.

Firstly, let's deal with Hamiltonian \eqref{AAmodelphi}.
By a gauge transformation $\ket{n}\rightarrow e^{-i\Phi n/N}\ket{n}$, Hamiltonian \eqref{AAmodelphi} becomes
\begin{eqnarray}\label{flux_trans}
H(\Phi)&=&\sum_{n} \Big[J_R e^{-i\Phi/N}\dyad{n+1}{n} +J_L e^{i\Phi/N}\dyad{n}{n+1})\notag\\
&&\quad+\Delta_n\dyad{n}{n}\Big].
\end{eqnarray}
Then, a Fourier transform, $\ket{n}=\frac{1}{\sqrt{N}}\sum_k e^{-i2\pi\beta kn}|k\rangle $, can further change it to the $k$-space,
\begin{eqnarray}\label{AAmodel1dual}
H{(\Phi)}=\sum_k\big[\Delta\big(\dyad{k+1}{k} +\dyad{k}{k+1}\big)+J_k(\Phi)\dyad{k}{k}\big],\notag\\
\end{eqnarray}
where $J_k(\Phi)=2J[\cosh\alpha\cos(2\pi\beta k+\Phi/N)-i\sinh\alpha\sin(2\pi\beta k+\Phi/N)]$. Note that the quasimomentum is $2\pi\beta k$, not the index $k$; The hopping term actually couple the two quasimomenta with difference $2\pi\beta$. Due to the PBCs, the quasimomentum should satisfy $2\pi\beta k=2\pi m/N$, i.e., $k=m/\beta N$, where $m\in$ integer.
To make the Hilbert space closed, we can just set $\beta=p/N$, and thus, $k+1=(m+p)/p$ corresponds to another quasimomentum index in the same Hilbert space, if considering the periodicity of the Brillouin zone. In this sense, the two dual models, Eqs. \eqref{flux_trans} and \eqref{AAmodel1dual}, are equivalent with identical energy spectra.

Secondly, consider the Hamiltonian \eqref{AAmodel1} under OBCs with infinite length, i.e., $N\rightarrow\infty$. The dual Hamiltonian in $k$-space has the same form as Eq. \eqref{AAmodel1dual} with only the difference that $\Phi=0$ and the boundaries are open.
When $J_R=J_L=J$, i.e., $\alpha=0$, the dual Hamiltonians have the same form and thus $\det h'(\Delta,J)=\det h'(J,\Delta)$, i.e., $J^N\det h'(\Delta/J)=\Delta^N\det h'(J/\Delta)$. Note that $\det h=\det h'$ because of their similarity, we have the relation that $\det h(\Delta/J)=(\Delta/J)^N \det h(J/\Delta)$.

We have noted that Ref. \cite{181017-8} \textit{numerically} gives the condition for the AL of the on-site complex AA model \eqref{AAmodel1dual}, $|J/\Delta\cdot\cosh\alpha|+|J/\Delta\cdot\sinh\alpha|=1$, i.e., $\Delta/J=e^{|\alpha|}$, which is consistent with our result in the main text.
\section{Calculation of the winding number}\label{winding}
We calculate the winding number \eqref{winding1} of Hamiltonian \eqref{AAmodelphi}. In matrix form, it can be rewritten as
\begin{equation}
\hat{H}_{\Phi}= \sum_{mn} h_{mn}(\Phi)\dyad{m}{n},
\end{equation}
where $h_{mn}(\Phi)$ is the entry of the following matrix,
\begin{equation}\label{htwist}
h(\Phi)=\begin{pmatrix}
\Delta_1&J_L&&&J_R e^{-i\Phi}\\
J_R&\Delta_2&J_L&&\\
&\ddots&\ddots&\ddots&\\
&&J_R&\Delta_{N-1}&J_L\\
J_L e^{i\Phi}&&&J_R&\Delta_N
\end{pmatrix}.
\end{equation}
The key to calculate the winding number is the determinant of $h(\Phi)$. Mathematically, we have
\begin{eqnarray}\label{det}
&&\det h(\Phi)=-(-J_L)^{N} e^{ i\Phi}-(-J_R)^{N} e^{-i\Phi}+P\notag\\
&&\quad=-2(-J)^N (\cosh \alpha N \cos \Phi+i\sinh\alpha N\sin \Phi)+P,\notag \\
\end{eqnarray}
where $P=\det h'-J^2\det u'$ with $h'$ being defined in Eq. \eqref{hprime} in the main text and $u'$ is a submatrix with $(N-2)$ dimension of $h'$ by removing the first and last row and column. Apparently, $P$ is real.

Because the winding number \eqref{winding1} reveals how det$\hat{H}(\Phi)$ evolves with respect to $\Phi$ from $2$ to $2\pi$ in the complex plain, we can rewrite the winding number with the aid of the sign operators% \cite{}, 
\begin{equation}
\nu=\frac{1}{2}\sum_i\text{sgn}[x(\Phi_i)]\cdot\text{sgn}\Big[\frac{dy(\Phi_i)}{d\Phi}\Big],
\end{equation}
where $x=\Re[\det h(\Phi)]=P-2(-J)^N \cosh \alpha N \cos \Phi$ and $y=\Im[\det h(\Phi)]=-2(-J)^N \sinh \alpha N \sin \Phi$. $\Phi_i$ is $i$th solution of $y(\Phi)=0$. Here are two solutions $\Phi_1=0$ and $\Phi_2=\pi$. Therefore, we have
\begin{eqnarray}
\nu &=& \frac{(-1)^N\mathrm{sgn}(\alpha)}{2} \big[\mathrm{sgn}(P+2(-J)^{N}\cosh \alpha N)\notag\\
&&-\mathrm{sgn}(P-2(-J)^{N}\cosh \alpha N)\big] \notag\\
&=&\mathrm{sgn}(\alpha)\theta(2J\cosh \alpha N-|P|).
\end{eqnarray}
The transition point is determined by
\begin{eqnarray}
|P|=2J\cosh \alpha N\approx Je^{|\alpha|N},
\end{eqnarray}
i.e.,
\begin{eqnarray}
\mathcal{P}\equiv \sqrt[N]{| P|}\approx Je^{|\alpha|},
\end{eqnarray}
where the squiggly equal sign is for the large $N$ limit.
To calculate $P$, we can expand it as
\begin{equation}\label{P}
P= \sum_{n=0}^{[N/2]} c_{N-2n}(-J)^{2n}(2\Delta)^{N-2n}+\text{Res.},
\end{equation}
with
\begin{equation}
c_{N-2n}=\sum^{L}_{\substack{j_s=j_s-1+2,\\  s=1,2,..n}}\prod^{N}_{\substack{i=1\\i\neq j_s, j_s+1,\\  (s=1,...n)}}\cos (2\pi\beta i),
\end{equation}
where $[N/2]$ means the nearest integer less than $N/2$, and ``Res.'' is the residual if 
$N/2$ is not an integer.

For the coefficient $c_N=\prod^{N}_{i=1}\cos(2\pi\beta i)$, we have
\begin{eqnarray}
\lim_{N \to \infty} \mathrm{In} c_N &=& \lim_{N \to \infty}\sum^{N}_{i=1}\mathrm{In}\cos(2\pi\beta i) \\
&=& N\int^{1}_{0}\ln\cos(2\pi\beta N x)dx \\
&=&-\frac{1}{2\pi\beta}\mathcal{L}(2\pi\beta N)\approx -N\mathrm{ln}2
\end{eqnarray}
where
\begin{eqnarray}
\mathcal{L}(x)&=& - \int^{x}_{0}\ln\cos(x')dx'\notag \\
&&= x\mathrm{ln}2-\frac{1}{2}\sum_{k=1}^{\infty}(-1)^{k-1} \frac{\sin(2kx)}{k^2}.
\end{eqnarray}
This means in the limit $N\rightarrow\infty$, $c_N\sim 2^{-N}$. In the same way, $c_{N-2}\sim 2^{-(N-2)}$.
Thus, using Eq. \eqref{P}, we have
\begin{equation}
\mathcal{P}=J \left[\left|c_N\left(\frac{2\Delta}{J}\right)^{N}-c_{N-2}\left(\frac{2\Delta}{J}\right)^{N-2}+...\right|\right]^{1/N}.
\end{equation}
For $\Delta/J\le 1$, $\lim_{N\to \infty}\mathcal{P}=J$, and thus $\nu = \mathrm{sgn}(\alpha)$, while for $\Delta/J>1$, $\lim_{N\to \infty}\mathcal{P}=\Delta$ and thus $\nu= \mathrm{sgn}(\alpha)\theta(Je^{|\alpha|}-\Delta)$, that is, when $e^{|\alpha|}<\Delta/J$, $\nu=0$ and when $e^{|\alpha|}>\Delta/J$, $\nu=\mathrm{sgn}(\alpha )$.

\section{Phase diagrams for other cases}\label{other}
In the main text, we paid attention to the typical case of positive $J_L$ and $J_R$ in Hamiltonian \eqref{AAmodel1}. Here we show that the general case is related to this special case, and thus share the same transition point on AL.

The Hamiltonian with arbitrary complex hoppings reads
\begin{eqnarray}\label{generalmodel}
\hat{H}_\text{gel}&=&\sum_n(J_Re^{i\phi_R}\dyad{n+1}{n} +J_Le^{i\phi_L}\dyad{n}{n+1}\notag \\
&&+\Delta_n \dyad{n}{n}),
\end{eqnarray}
where $J_{R(L)}>0$ and $\Delta_n$ keep the same definitions as in Hamiltonian \eqref{AAmodel1} of the main text, and $\phi_{R(L)}$ is the arbitrary argument of the corresponding hopping.
To reveal the relation between the general case of hoppings and the positive case, we do the following gauge transformation, which does not change the energy spectrum,
\begin{eqnarray}\label{genU}
\hat{U}\hat{H}_\text{gel}\hat{U}^{-1}&=&e^{i\frac{\phi_R+\phi_L}{2}}\sum_n\Big(\Delta_ne^{-i\frac{\phi_R+\phi_L}{2}}\dyad{n}{n}\notag\\
&&+J_R\dyad{n+1}{n} +J_L\dyad{n}{n+1}\Big),
\end{eqnarray}
where $\hat{U}$ is a unitary operator defined by $\hat{U}\ket{n}=e^{i\frac{\phi_L-\phi_R}{2}n}\ket{n}$.
Except for the overall phase and the phase of on-site terms, the above transformed Hamiltonian is similar to Hamiltonian \eqref{AAmodel1}.

Specifically, when $\phi_R+\phi_L=2n\pi$ (n $\in$ integer), we have
\begin{eqnarray}
\hat{H}_\text{gel}&=&(-1)^n\hat{U}^{-1}\hat{H}\hat{U}
\end{eqnarray}
where $\hat{H}$ is just the Hamiltonian \eqref{AAmodel1} in the main text.
Apparently, the phase boundaries of this case is identical to the real-hopping case with only the eigenenergy $E$ becoming $(-1)^nE$.
Note that for odd $n$, the minus sign of on-site terms in Eq. \eqref{genU} can be absorbed to the cosine terms in $\Delta_n$ by shifting a phase, which makes no difference for the infinite chain.

\begin{figure}[tbp]
	\centering
	\includegraphics[width=1\columnwidth]{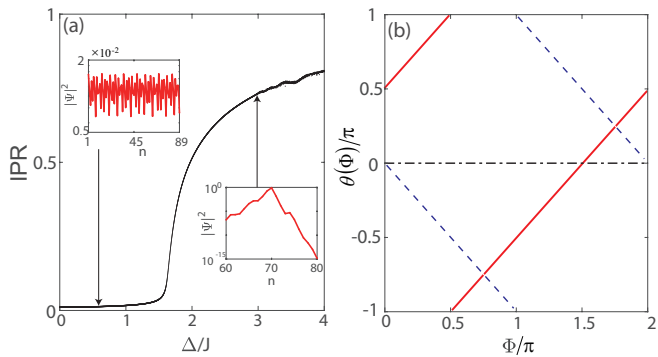}
	\caption{Phase transition for $(\phi_R,\phi_L)=(0,\pi/2)$. 
		(a) $\overline{\text{IPR}}$ v.s. $\Delta/J$ for $\alpha=0.5$ under PBCs.
		Insets: The profiles of the eigenstates of $10$th lowest $|E|$, showing the NHSE and the AL at $\Delta/J=0.5$ and $3$, respectively. 
		(b) $\theta(\Phi)$ for $\alpha=\Delta/J=0.5$ (solid red), $\alpha=-\Delta/J=-0.5$ (dashed blue), and $\alpha=0,\Delta/J=3$ (dash-dotted black), which correspond to $\nu=+1$, $-1$, and $0$, respectively.
		The calculation is carried on with $N=89$ and $\beta=55/89$.
	}\label{gelphase}
\end{figure}

For the general case, we cannot find a relation to the positive real-hopping case, which can be understood by noting that the right and left hoppings generally generate a net flux, $\phi_L+\phi_R$, for each two nearest-neighbor sites, as there seems a coil inbetween with a magnetic field through it, and thus, the phase cannot be gauged away.
However, the phase diagrams seems the same by our numerical calculation, which can also be characterized by the winding number, as shown in Fig. \ref{gelphase}. 

\end{document}